\documentstyle[aps,prl,epsf,multicol]{revtex}
\begin{document}
\draft
\title{Finite-Size Effects on Critical Diffusion 
and Relaxation Towards Metastable Equilibrium}
\author{W. Koch and V. Dohm}
\address{Institut f\"ur Theoretische Physik, Technische Hochschule Aachen, D-52056 Aachen, Germany}
\date{\today}
\maketitle
\begin{abstract}
We present the first analytic study of finite-size effects on 
critical diffusion 
above and below $T_c$ of three-dimensional 
Ising-like systems whose order parameter
is coupled to a conserved density.
We also calculate the finite-size relaxation time that
governs the  critical order-parameter relaxation
towards a metastable equilibrium state below $T_c$. 
Two new universal dynamic amplitude ratios at $T_c$ are predicted and
quantitative predictions of dynamic finite-size scaling
functions are given that can be tested by Monte-Carlo simulations. 
\end{abstract}
\begin{figure}
\narrowtext
\epsfxsize=\hsize\epsfbox{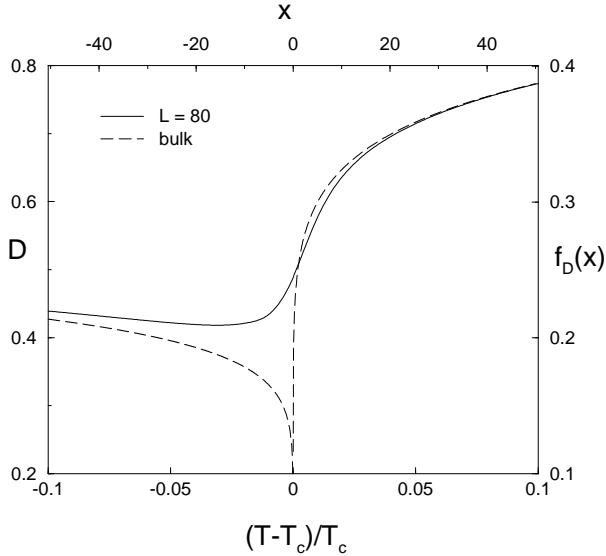}
\caption{Diffusion constant 
$D(\tilde{t}, L)/A^+_D$ for $L = 80 \xi_0$ (solid line)
vs $\tilde{t}$,
and of the scaling function $f_D(x)$, Eq.~(\ref{fdx}), vs $x = \tilde{t} 
 L^{(1-\alpha)/\nu}$ (solid line), with $L$ in units of $\xi_0$.
The dashed lines represent the bulk diffusion constants 
$D^\pm (\tilde{t})/A^+_D$. \label{fig1}}
\end{figure}

\begin{multicols}{2}
\narrowtext
The dissipative critical dynamics of {\it bulk} systems with a 
non-conserved order parameter 
are fairly well understood. Depending on
whether the order parameter is governed by purely relaxational dynamics
or whether it is coupled to a hydrodynamic (conserved) density such 
systems belong to the universality classes of models A or  C \cite{hoha,hhm}.
The fundamental dynamic quantities of these systems are the relaxation and
diffusion times which diverge as the critical 
temperature $T_c$ is approached. \\
\indent For {\it finite} systems, these times are
expected to become smooth and finite throughout the
critical region and to depend sensitively on the geometry and boundary
conditions.  These finite-size effects are particularly large in Monte Carlo
(MC) simulations of small systems.
On a qualitative level, they can be interpreted 
on the basis of phenomenological
finite-size scaling assumptions. 
For a more stringent analysis the knowledge
of the shape of universal finite-size scaling functions is necessary. 
So far there exist reliable theoretical predictions on finite-size dynamics
in three dimensions only on two relaxation times $\tau_1$ and 
$\tau_2$ determining the
long-time behavior of the order parameter and the square of the order 
parameter  \onlinecite{kds,gonidi}. No 
analytic work exists, to the best of our knowledge, on the important 
universality class \cite{hoha} of diffusive finite-size behavior near $T_c$.
This is of relevance, e.g.,  to magnetic systems with mobile
impurities \cite{hoha},  to binary alloys with an order-disorder transition 
\cite{juel,eis},
to uniaxial antiferromagnets \cite{hoha}
or to systems whose order parameter is coupled
to the conserved energy density \cite{hoha,hhm}. \\
\indent In this Letter we present the
first  prediction
of the finite-size scaling function for the critical diffusion time of
three-dimensional systems  above and below $T_c$. Furthermore we shall 
present the analytic identification and quantitative calculation of
a new leading relaxation time that governs the
critical order-parameter relaxation towards a {\it metastable} equilibrium
state  of finite  systems below $T_c$. 
Our predictions
contain no adjustable parameters other than two amplitudes of the bulk system.
\\
\indent We start from model C \cite{hhm}, i.e., from the  
relaxational and diffusive Langevin equations for the 
one-component order-parameter 
field $\varphi({\bf x},t)$ and for the density
$\rho({\bf x}, t) \; = \; \langle\rho\rangle + \; m ({\bf x},t)$ in a finite
volume $V$, 
\begin{eqnarray}
&& \frac{\partial \varphi({\bf x}, t) }{\partial t} =  
- \Gamma_0 \, \,  \frac
{\delta H }{\delta \varphi({\bf x},t)} \, \, 
+ \, \, \Theta_\varphi ({\bf {x}}, t)
 , \label{lan} \\
&& \frac{\partial m({\bf x}, t) }{\partial t}  = 
\lambda_0 \, \nabla^2  \, \,   \frac
{\delta H }{\delta m ({\bf x},t)} \, \, 
+ \, \, \Theta_m ({\bf {x}}, t) \; , \\
&& H   =  \mbox{$\int_V$} d^{d}x \, 
\big [ \mbox{ $\frac{1}{2}$}  \tau_0   \varphi^2 
+ \mbox{$\frac{1}{2}$}  (\nabla \varphi)^2 + \tilde{u}_0 
\varphi^4 \nonumber \\
 && \qquad \qquad \qquad+  \mbox{$\frac{1}{2}$} m^2  +  \gamma_0 m \varphi^2 
 -  h_0 m  \big ] \label{ham} 
\end{eqnarray}
where $\Theta_\varphi$ and $\Theta_m$ are Gaussian 
$\delta$-correlated random forces.   
We consider an equilibrium ensemble near $T_c (\bar{\rho})$ 
at fixed 
$\bar{\rho} = V^{-1} \int_{V} d^dx\; \rho ({\bf x},t)$. 
This corresponds to the
experimental situation of keeping the conserved quantity (e.g., number
of impurities) fixed when changing the reduced temperature 
$\tilde{t} = 
\big [ T-T_c(\bar{\rho}) \big ]/T_c(\bar{\rho})$. The latter
enters through $\tau_0$. 
Because of $\langle\rho\rangle = \bar{\rho}$ we have $ \langle m \rangle = \bar{m} = 0$.
Eqs.~(\ref{lan})~-~(\ref{ham}) describe the dynamics of  relaxational and 
diffusive modes that are coupled through $\gamma_0$. We are interested
in the long-time behavior 
of the diffusive modes above, at and
below $T_c$ as well as in the order-parameter relaxation on an intermediate
time scale below $T_c$. We shall begin with the diffusive modes.  
For simplicity and for the purpose of a comparison with 
MC simulations, we assume
cubic geometry, $V = L^d$, with periodic boundary conditions. \\
\indent For the {\it bulk} system, the diffusion constants 
$D^\pm(\tilde{t}) $ above and below $T_c$
appear in the small $k$ limit of the long-time
behavior of the correlation function 
\begin{equation}
C_n (k,\tilde{t}, t) =  V^{-1} \langle  n_{\bf k}(t) n_{-{\bf k}} (0) \rangle 
\sim \exp \left [ - D^\pm(\tilde{t}) k^2 t \right ] 
\end{equation}
where 
$n_{\bf k}(t) = m_{\bf k} (t) + c_n (k) \psi_{\bf k}(t)$
is an appropriate linear combination of  
$m_{\bf k}(t) 
= \int_V d^d x \; m ({\bf x},t) 
e^{-i{\bf k x}} $ and $ \psi_{\bf k}(t) =  \int_V d^dx \left 
[ \varphi ({\bf x},t) -  \langle\varphi\rangle \right ] e^{-i{\bf kx}}$ 
with $c_n(k) = \tilde{c}_n k^2 + O(k^4)$.
The coefficient $c_n$ can be identified by linearizing Eqs.~(\ref{lan})~-~(\ref{ham}).
Above $T_c$, $c_n = 0$ because of $\langle\varphi\rangle = 0$. At
$T_c$, the long-time behavior of $C_n$ is non-exponential (power law)
for the bulk system. \\
\indent For the {\it finite} system, the coefficient $c_n(k)$ 
is modified [via the replacement $\langle\varphi\rangle \rightarrow M_0$ as
defined in Eqs. (\ref{m0}) and (\ref{cnk}) below] and the long-time behavior 
of $C_n$ 
remains exponential, 
\begin{equation}
C_n(k, \tilde{t}, L, t) \; \sim \; \exp \left [ - \Omega_n (k,\tilde{t},L) \; t \right ] \; , \label{l-cor}
\end{equation}
even in the non-hydrodynamic region
at bulk $T_c$ where the small-$k$ approximation is no longer justified.
As a con\-ceptual complication there exists a smallest {\it nonzero}
value $k^2_{min}  = 4 \pi^2/L^2$ of $k^2$ which prevents 
us to perform the 
limit $k \rightarrow 0$ for the finite system. Therefore we need to 
derive the finite-size scaling function for 
$\Omega_n (k, \tilde{t}, L)$
at finite $k$. Nevertheless we may define an effective diffusion time 
$\tau_D  = \Omega_n  (2 \pi L^{-1}, \tilde{t},L)^{-1}$ or a diffusion constant 
$D = \Omega_n/k^2$ at $k = k_{min} = 2 \pi/L$ 
of the finite system by 
\begin{equation}
D(\tilde{t},L) \; = \; (2 \pi )^{-2} L^2  \; \Omega_n (2 \pi L^{-1}, \tilde{t}, L)  \label{dtl}
\end{equation}
which interpolates smoothly between the bulk result 
$D(\tilde{t}, \infty) = D^\pm(\tilde{t}) $ above and below $T_c$ 
(Fig.~\ref{fig1}).

\begin{figure}
\narrowtext
\epsfxsize=\hsize\epsfbox{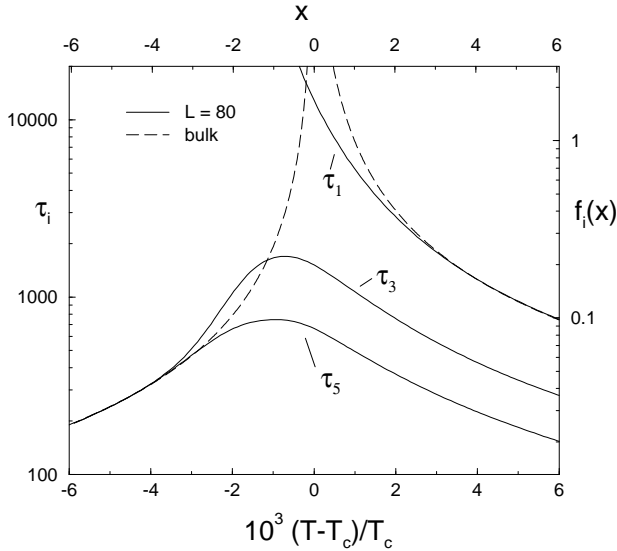}
\caption{Relaxation times
$\tau_i(\tilde{t}, L)/A_{\tau b}^+$ for $L = 80 \tilde{a}$ (solid lines)
vs $\; \tilde{t} = (T-T_c)/T_c$, and of their scaling functions
$f_i(x) \;$ vs $\; x = \tilde{t} \; L^{1/\nu}$ (solid lines),
with $L$ in units of the lattice constant $\tilde{a}$. Dashed lines:
bulk relaxation times $\tau^\pm_b (\tilde{t})/A^+_{\tau b}$.
\label{fig2} }
\end{figure}

In the spirit of finite-size theory \cite{kds,edc} we decompose
$\varphi ({\bf x},t) = M_0 +  \delta \varphi ({\bf x}, t) $
with the zero-mode average 
\begin{equation}
M^2_0 =  \int^\infty_{-\infty} dM \, M^2  e^{-H_0} \big / \int^\infty_{-\infty} dM  \, e^{-H_0} \label{m0}
\end{equation}
where $H_0 (M) = L^d (\mbox{$\frac{1}{2}$}
\tau_0 M^2 + \tilde{u}_0 M^4)$ is the $k=0$ part of 
$H$, with $M = V^{-1} \int_V d^dx \; \varphi$. For the finite
system, the quantity $M_0 (\tau_0,L)$ is non-zero for all $T$ and 
interpolates smoothly
between $T > T_c$ and $T < T_c$.
Linearization of Eqs. (\ref{lan}) - (\ref{ham}) with respect to $\delta \varphi_{\bf k}(t)$
and $m_{\bf k}(t)$ leads to 
\begin{eqnarray}
&& c_n(k)  = (w_0 \tilde{\gamma}_0)^{-1} \big \{ b^-_0  -  
\left [ (b^-_0)^2  +  w_0 \tilde{\gamma}_0^2 k^2   \right ]^{\frac{1}{2}}
 \big \}\; , \label{cnk} \\
&& \Omega_n (k, \tilde{t}, L) =  \mbox{$\frac{1}{2}$} \lambda_0 
\big \{ b^+_0  -  \left [ (b^-_0)^2  
+  w_0 \tilde{\gamma}_0^2 k^2 \right ]^{\frac{1}{2}} \big \} \; , \\
&& b^\pm_0(k) =  w_0 \, ( \tau_0 +  12 \tilde{u}_0 M^2_0 +  k^2)
 \pm   k^2 , \label{bpm}
\end{eqnarray}
with $w_0 = \Gamma_0/\lambda_0$ and $\tilde{\gamma}_0 = 4 \gamma_0 M_0$. \\
\indent An application of these unrenormalized expressions to 
the critical region requires us to turn to the renormalized theory.
The strategy of the field-theoretic renormalization-group
(RG) approach at $d = 3$ dimensions is
well established in bulk statics \cite{do1} and dynamics \cite{do2} and has 
been successfully applied recently to the model-A finite-size dynamics \cite{kds}.
The details of its application to model $C$ will be given elsewhere \cite{kd}.
Here we only present the asymptotic finite-size scaling form
\begin{equation}
\Omega_n(k, \tilde{t}, L) \; = \; L^{-z} f_n 
(\tilde{t}L^{(1-\alpha)/\nu}, k L) 
\end{equation}
as derived from Eqs.~(\ref{l-cor}) and (\ref{cnk})~-~(\ref{bpm}), with the 
dynamic critical exponent
$ z = 2 + \alpha/\nu$ \cite{hhm}. The scaling function reads
in three dimensions
\begin{eqnarray*}
f_n(x,\kappa) & = & A_n \; \tilde{\ell}^{\alpha/\nu} 
\big \{ b_+ \; - \; 
\big [ b_-^2 \; + \;w^* c^* \tilde{\ell}^{1/2} \kappa^2 \vartheta_2
(\tilde{y}) \big ]^{\frac{1}{2}} \big \} \; , \\
b_\pm (x,\kappa) & = & w^* [ \tilde{\ell} (x)^2 \; + \; \kappa^2 ]
\pm \kappa^2 \; , \\
\tilde{\ell}(x)^{3/2} & =  & (4\pi \tilde{u}^*)^{1/2} \big \{ \tilde{y} (x) 
\; + \; 12 \vartheta_2 (\tilde{y}(x)) \big \} \; , \\
\tilde{y}(x) &   =  & (4\pi \tilde{u}^*)^{-1/2} \tilde{\ell}
(x)^{(3/2)-(1-\alpha)/\nu}
\; \hat{x} \; , \\
\vartheta_2 (y) & =  & (\mbox{$ \int^\infty_0 $} ds \;  s^2 
e^{-\frac{1}{2}ys^2-s^4}  ) \big / ( \mbox{$ \int^\infty_0 $} ds  
\; e^{-\frac{1}{2} y s^2-s^4}  ),
\end{eqnarray*}
with $c^* \;  = \; 16(\gamma^*)^2 (4\pi/\tilde{u}^*)^{1/2}$ 
and  $\hat{x} \; = \; x \;  \xi_0^{-(1-\alpha)/\nu}$.
This yields the scaling form $D(\tilde{t},L) \; = \; 
L^{2-z} f_D(x)$
for the diffusion constant, Eq.~(\ref{dtl}), with 
\begin{equation}
f_D(x)  =  (2\pi)^{-2} f_n (x, 2 \pi).  \label{fdx}
\end{equation}
The static parameters are \cite{do1} $\tilde{u}^* = u^* + (\gamma^*)^2/2$ and
$(\gamma^*)^2 = \alpha [ 4 \nu B (u^*) ]^{-1}$
with \cite{lmsd} $u^* = 0.0404$
and $B(u^*) = 0.502$ in three dimensions. For $\nu$ and $\alpha$
we take $0.6335$ and $0.100$ \cite{liu}. The dynamic parameter is $w^* = 1$ 
in one-loop order.
The two non-universal bulk amplitudes $\xi_0(\bar{\rho})$
and $A_n = \mbox{$\frac{1}{2}$} A^+_D  \xi_0^{z-2}$
are defined by the asymptotic behavior $\xi =  \xi_0  
\tilde{t}^{-\nu/(1-\alpha)}$ and $D^+  (\tilde{t}) =  
A^+_D \tilde{t}^{(z-2)\nu/(1 - \alpha )}$
of the correlation length and diffusion constant at 
fixed $\bar{\rho}$ above $T_c$.
The exponent $\nu/(1-\alpha)$
instead of $\nu$ is due to Fisher renormalization \cite{fis}. \\
\indent The solid line in Fig.~\ref{fig1} shows $D(\tilde{t},L)/A^+_D$ vs $\tilde{t}$
for the example $L = 80 \xi_0$. 
The same line represents $f_D(x)$ vs $x$ (top scale).
For comparison the bulk limits $D^\pm$
(dashed lines) are also shown, with 
$A^-_D/A^+_D = 2^{\alpha/(1-\alpha)} (1 + \mbox{$\frac{1}{2}$} 
\gamma^{*2}/u^*)^{-1} = 0.55$. We expect the accuracy of these results to
be of $O(10 \%)$. These predictions can be tested
by MC simulations, after adjusting $\xi_0(\bar{\rho})$ and $A^+_D$ in the 
bulk region $x \gg 1$ above $T_c$. \\
\indent In addition to the finite-size effect on the diffusive modes there 
exists an interesting finite-size effect on the relaxational modes
below $T_c$ that has so far not been investigated analytically. 
It is well known
that no spontaneous 
symmetry breaking can take place in 
finite systems below $T_c$ because of ergodicity. For
Ising-like systems, ergodicity
implies
a "tunneling" between metastable states of opposite orientation
of the magnetization  as observed in MC simulations \onlinecite{sto,heu,wl}.
On an intermediate time scale $t < t_x(L)$, however, the magnetization 
does not
change sign and its magnitude relaxes towards a {\it finite} value that 
characterizes such a metastable state \cite{sto}. 
This relaxation process is important for large systems since the 
crossover time $t_x(L)$ is expected to grow with the size $L$ as
$\sim L^z$ where $z$ is the dynamic critical exponent. This process occurs
both in model A and model C, therefore we confine ourselves to the 
simpler model A in the following. We stress that the relaxation process
for $t < t_x(L)$ is
fundamentally different from the ultimate long-time behavior 
for $t \gg t_x(L)$
studied previously \onlinecite{kds,gonidi}. \\
\indent Model A is defined by Eq.~(\ref{lan})
where $H$ is replaced by 
\begin{equation}
H_\varphi   =   \int_V d^{d}x \, 
\left [ \mbox{ $\frac{1}{2}$} \,  r_0 \,  \varphi^2 
+ \mbox{$\frac{1}{2}$} \, (\nabla \varphi)^2 + u_0 
\, \varphi^4  \right ] \; . 
\end{equation}
We consider the time-dependent spatial average  
$M(t,L) = L^{-d}  \int_V d^dx \varphi ({\bf x},t) \; .$ 
We are primarily interested in the long-time behavior of the equilibrium
correlation function $\langle M(t,L) M(0,L)\rangle \equiv C(t,L)$ for $d = 3$.
For the bulk system, this behavior is
\begin{eqnarray}
C (t,\infty)  & \sim &  A^+_{b} \exp (-t/\tau^+_b) ,\\
C (t,\infty)   -      M^2_{sp} & \sim & A^-_{b} \exp (-t/\tau^-_b)
\label{expo}
\end{eqnarray}
above and below $T_c$ where 
$M_{sp} = \lim_{t\rightarrow \infty} \langle M(t, \infty) \rangle$ 
is the spontaneous order parameter and 
$\tau^\pm_b$ are the bulk relaxation times. 
For the finite system
{\it above} $T_c$, the leading time dependence is still a single
exponential $\sim c_1 e^{-t/\tau_1(L)}$ 
with a relaxation time
$\tau_1(L)$ whose finite-size scaling function is known both analytically
\onlinecite{kds,gonidi}
and numerically \cite{kds,wl}. In particular, 
$\lim_{L \rightarrow \infty} \tau_1(L) = \tau^+_b$. \\
\indent For the finite system {\it below} $T_c$, however, the situation is more
complicated and considerably less well explored. MC simulations \cite{sto} 
and
phenomenological  considerations suggest that there should exist an $L$
dependent generalization 
$\tau^-(L)$ of $\tau^-_b$ , with $\lim_{L \rightarrow \infty} \tau^-(L) 
= \tau^-_b$, which should describe (i) the exponential relaxation of 
$ \langle M(t,L) \rangle $ towards a metastable finite value on an 
intermediate time
scale $t < t_x (L)$ before tunneling sets in, and (ii)
a corresponding exponential decay of $C(t,L)$ on this time scale. 
The question arises whether and how this important relaxation time
$\tau^-(L)$ can be identified analytically within models A and C. 
This question was left
unanswered in the previous literature. In particular, neither $\tau_1(L)$
nor $\tau_2(L)$ as calculated previously \cite{kds}, can be identified
with $\tau^-(L)$. [Below $T_c, \tau_1$ describes the  
decay of $\langle M(t,L) \rangle$ and of $C(t,L)$ towards
{\it zero} for $t \gg t_x(L)$ due to tunneling processes, and $\tau_2$
describes the decay of $\langle M(t)^2 \rangle$ and of  
$\langle M(t)^2 M(0)^2 \rangle$ towards $\langle M^2 \rangle_{eq}$ and
$\langle M^2 \rangle_{eq}^2$, respectively, for
$t \gg t_x (L)$.] In the
following we establish an analytic identification of $\tau^-(L)$ and
present the first quantitative prediction for its finite-size scaling behavior.
\\
\indent To elucidate the main features we first neglect the 
inhomogeneous fluctuations 
$\sigma ({\bf x},t) = \varphi ({\bf x},t) - M(t)$.
Then Eq.~(\ref{lan}) is equivalent to the Fokker-Planck equation 
$\partial P (M,t)/\partial t = -{\cal{L}}_0 P(M,t)$
for the probability distribution $P(M,t)$ with the  operator
\begin{equation}
 {\cal L}_0  = - \frac{\Gamma_0}{L^d} \, \, 
 \frac{\partial }{\partial M} \Big (  \frac{dH_0(M)}{dM} + 
\frac{\partial}{\partial M} \Big ) \label{oper}
\end{equation}
where
$H_0(M) = L^d (\mbox{$\frac{1}{2}$}
r_0 M^2 + u_0 M^4)$ . 
It is well known \cite{dvk} that $C(t,L)$
is determined by the eigenvalues $\epsilon_k$ and eigenfunctions  
$\phi_k (M)$ of ${\cal L}_0$ according to 
\begin{equation}
C(t,L)  =  \sum_{k = 1}^{\infty}   c_{k}(L) 
\exp [-t/\tau_k(L)], \; t > 0   \label{sum}
\end{equation}
with $c_k (L) = [ \int^\infty_{-\infty} d M \, M  \, \phi_k (M) ]^2 $
and $\tau_k (L) = \epsilon_k^{-1}$, 
$\epsilon_0 = 0 \leq \epsilon_1 \leq \epsilon_2 \dots$ 
By symmetry, $c_k = 0$ for even values of $k$.
Below $T_c$,
$\tau_1(L)$ diverges in the bulk limit and 
$\lim_{L \rightarrow \infty} c_1 e^{-t/\tau_1} = M_{sp}^2$ 
becomes time-independent,
thus an analysis of the
$k = 3$ term in Eq.~(\ref{sum}) becomes indispensable. From the spectrum
of ${\cal{L}}_0$ \cite{dvk} we find a degeneracy for $k = 3$
and $k = 5$ in the bulk limit  for $r_0 < 0$. This requires to
take the $k = 5$ term into account as well. We have found, however,
that the coefficient $c_5$ vanishes in the bulk limit below $T_c$ whereas
$c_3$ remains finite. For finite $L$ near $T_c$, $\tau_5$ is well 
separated from
$\tau_3$, as shown below. Thus it suffices to describe  
the time dependence of 
$C(t,L)$ on intermediate time scales $t \sim O(\tau^-(L))$
and $O(\tau^-(L)) < t < O(\tau_1 (L))$ as well as on the 
long-time scale $t \gg \tau_1 (L)$
as
\begin{equation}
C(t,L) \sim c_1(L)  e^{-t/\tau_1(L)} \;
+ \; c_3(L) e^{-t/\tau_3(L)}   \label{c1c3}
\end{equation}
where $c_1(\infty) = M^2_{sp}$
and  $c_3 (\infty) \;  =  \;  A^-_{b} $ below $T_c$. 
In particular we arrive at the desired identification
\begin{equation}
\tau^-(L) \equiv \tau_3(L) \; , \; \lim_{L \rightarrow \infty} \tau_3 (L)
\; = \; \tau^-_b \; . 
\end{equation}
We conclude that, although $\tau_3 (L)$ represents only
a subleading relaxation time above $T_c$, $\tau_3 (L)$ {\it governs the 
leading time dependence of} $C(t,L)$ {\it of 
large finite systems below} $T_c$ (Fig.~\ref{fig2}). \\
\indent These results yield also the key to the \mbox{interpretation} of
$\tau_3(L)$ as the  relaxation time governing the approach of the
{\it non-equilibrium} quantity $\langle M(t,L)\rangle$
towards a {\it metastable} finite value before 
$M(t,L)$ starts to change sign.
This interpretation is based on the fact \cite{tom} that~the leading 
relaxation times
of  $\langle M (t,L) \rangle$  are  determined by the same eigenvalues of  
${\cal{L}}_0$ as the long-time behavior of the  
equilibrium correlation function $C$, i.e., 
\begin{equation}
\langle M(t,L)\rangle \; \sim \; \tilde{c}_1 (L) \; e^{-t/\tau_1(L)} \; + \;
\tilde{c}_3 (L) \; e^{-t/\tau_3(L)} . \label{mtl}
\end{equation}
The basic difference between Eqs. (\ref{mtl}) and (\ref{c1c3}) is that the 
coefficients $\tilde{c}_k$ depend on the initial (non-equilibrium) state. \\
\indent We proceed by presenting the results of a quantitative calculation of
$\tau_3(L)$ and $\tau_5(L)$ including the effect of the inhomogeneous
fluctuations $\sigma({\bf x})$ to one-loop order. This calculation is 
parallel to that performed previously \cite{kds} and is expected
to be as reliable as the previous results \cite{kds}. 
It is based on the Fokker-Planck equation
$\partial P(M, t)/\partial t = - {\cal{L}}_1 P(M,t)$ where ${\cal{L}}_1$
has the same structure as  ${\cal{L}}_0$, Eq.~(\ref{oper}), but with 
$r_0,u_0,\Gamma_0$ replaced by (positive) effective parameters $r_0^{eff},
u_0^{eff}, \Gamma_0^{eff}$ \cite{kds,edc,repl}.
In terms of the eigenvalues $\mu_3(\kappa)$ and $\mu_5(\kappa)$ 
of the equivalent
Schr\"odinger equation \cite{dvk} we determine the relaxation times 
$\tau_3$ and $\tau_5$ as
\begin{equation} 
\tau_i  =    ( 2 \Gamma_0^{eff}  )^{-1} L^{d/2}
( u_0^{eff} )^{-1/2} \;  \mu_i(\kappa) 
\end{equation}
with
$\kappa  = \mbox{$\frac{1}{2}$} r_0^{eff} 
L^{d/2} (u_0^{eff})^{-1/2}$.
In the asymptotic  region 
the field-theoretic RG approach at $d = 3$ \onlinecite{kds,edc,do1,do2} yields the finite-size scaling form
$\tau_i = L^z f_i(x), i = 3,5$, with the scaling variable 
$x = \tilde{t} L^{1/\nu}$, $\tilde{t}=(T-T_c)/T_c$.
The analytic expressions for $f_i(x)$ are analogous to those given 
previously \cite{kds} and will be given elsewhere \cite{kd}. At $T_c$ we predict the universal ratios $\tau_1/\tau_3 = 8.5$ and $\tau_3/\tau_5 = 2.3$.

The results are shown in Fig.~\ref{fig2}. 
For an application to the Ising model
we have taken $\xi_0/\tilde{a} = 0.495$ \cite{liu} where $\tilde{a}$ is the
lattice spacing. The relaxation times $\tau_i$ in Fig.~\ref{fig2} are 
normalized to the bulk amplitude 
$A^+_{\tau b}$ of $\tau^+_b = A^+_{\tau b} \tilde{t}^{-\nu z} 
, \; z = 2.04$
(dashed line above $T_c$). 
Below $T_c$, our theory yields the expected \cite{hus,gra} exponential 
decay, Eq.~(\ref{expo}), for the $d=3$ bulk system, in disagreement with 
Ref.~\cite{tanami}.
The dashed line below $T_c$ represents
the bulk relaxation time $\tau^-_b = A^-_{\tau b} |\tilde{t}|^{-\nu z}$
with 
$A^-_{\tau b}/A^+_{\tau b} =  2^{-\nu z} 
(1 + \frac{9}{4} u^*)/(1 + 18 u^*) = 0.26$
in three dimensions.
Unlike for $\tau_1$ and $\tau_2$ \cite{kds,heu,wl}, 
no MC data are presently available for
$\tau_3$. \\
\indent In summary we have presented the first quantitative predictions for the
finite-size effects on  critical diffusion and
order-parameter relaxation towards metastable equilibrium in
three-dimensional systems near $T_c$.
It would be interesting to test the predicted universal ratios 
$\tau_1/\tau_3$, $\tau_3/\tau_5$ and the finite-size scaling functions
$f_D(x)$ and $f_i(x)$  (Figs. 1 and 2) by MC simulations. This
appears to be within reach of present simulation techniques \cite{sen}. \\
\indent Support by SFB 341 der Deutschen Forschungsgemeinschaft
and by NASA is acknowledged.

\end{multicols}

\begin{references}
\bibitem{repl} In Eqs. (7) and (11) of Ref. 3, $\Gamma_0$ should be replaced by $\Gamma_0^{eff} = \Gamma_0 [1 + 144 u^2_0 M_0^2 S_3(r_{0L})]^{-1}$,
see also Ref. 10. This has a negligible effect on $f_1(x)$ and $f_2(x)$.
\end{references}
\end{document}